\documentclass[11pt,onecolumn,amssymb,nofootinbib]{revtex4}
\usepackage{amsmath, amsthm, amscd, amssymb}
\usepackage{bm}
\usepackage{bbm}

\begin{document}

\title{\bf $SU(n)$ symmetry breaking by rank three and rank two antisymmetric tensor scalars}

\author{Stephen L. Adler}
\email{adler@ias.edu} \affiliation{Institute for Advanced Study,
Einstein Drive, Princeton, NJ 08540, USA.}

\begin{abstract}
We study $SU(n)$ symmetry breaking by rank three and rank two antisymmetric tensor fields.
Using tensor analysis, we derive branching rules for the adjoint and antisymmetric tensor
representations, and explain why for general $SU(n)$ one finds the same $U(1)$ generator mismatch
that we noted earlier in special cases.  We then compute the masses of the various scalar fields in
the branching expansion, in terms of parameters of the general renormalizable  potential for the
antisymmetric tensor fields.

\end{abstract}

\maketitle

\section{Introduction}
The most familiar case of symmetry breaking for grand unified theories, such as
minimal $SU(5)\supset SU(2) \times SU(3) \times U(1)$, utilizes a scalar field in the adjoint representation, with
a gauge singlet component with $U(1)$ generator zero that receives a vacuum expectation.
The symmetry breaking mechanism is then  straightforward: since the gauge fields and the
symmetry breaking scalar are both in the adjoint representation, the same representations
appear in their branching expansions. As a consequence, the massless gauge fields that pick up masses,
and the scalars that supply their longitudinal components, have the same group theoretic quantum numbers.

We recently noted \cite{adler1, adler2} that when the symmetry breaking scalar is in a totally
antisymmetric representation, the situation is more complicated.  Using as explicit examples
$SU(8)$ broken by a  rank three antisymmetric tensor scalar, and $SU(5)$ broken by a rank two
antisymmetric tensor scalar, we showed that there is a mismatch between the $U(1)$ generator
values of the massless gauge fields that obtain masses, and the scalars that supply their
longitudinal components.  We noted that this mismatch is related to the fact
that the gauge singlet component of the antisymmetric tensor field that receives a vacuum
expectation has a nonzero $U(1)$ generator $N$, requiring a {\it modular} ground state that
is periodic in integer divisors $p$ of $N$.

The purpose of this paper is twofold.  First, we show that the mismatch found in \cite{adler1,adler2} appears
in the case of general $SU(n)$, and can be traced to the fact that  invariant tensors
lying in $SU(3)$ or $SU(2)$ subgroups are available to lower subgroup indices.  This analysis is given
in Sec. 2, where we use tensor methods to compute the relevant branching expansions and $U(1)$ generator
values.  The second aim of this paper is to calculate the masses of the various scalar field components
in the branching expansions, obtained by expanding the general renormalizable scalar field potential
around the generic symmetry breaking minimum.  This analysis is given in Sec. 3, and a brief summary
of our results follows in Sec. 4.

Our notation is to define  the  upper index totally antisymmetric tensor with $R$ components to be a basis for the representation $R$, and the corresponding lower index tensor
to be a basis for the conjugate representation $\overline{R}$. Thus in $SU(n)$ the tensor $\phi^\alpha,\,\alpha=1,...,n$ is a
basis for the fundamental representation $n$, and $\phi_{\alpha}$ is a basis for the conjugate representation $\overline{n}$.
In $SU(3)$, the tensor $\phi^{\alpha}$ is a basis for the 3, and since the totally antisymmetric tensor $\epsilon_{\alpha \beta \gamma}$
is invariant and can be used to lower indices, both the tensors $\phi_{\alpha}$ and $\phi^{[\alpha\beta]}$  give a basis for
the $\overline{3}$, and the tensor $\phi^{[\alpha\beta\gamma]}\propto \epsilon^{\alpha \beta \gamma}$ is a singlet.    Similarly, in $SU(2)$, since the invariant tensor $\epsilon_{\alpha \beta}$ can be used to lower indices
the representations 2 and $\overline{2}$ are equivalent, and can be represented by either $\phi^{\alpha}$ or $\phi_{\alpha}$, and the
tensor $\phi^{[\alpha\beta]}\propto \epsilon^{\alpha \beta}$ is a singlet \cite{coleman}.

\section{Branching rules for the  $SU(n)$  antisymmetric tensor and adjoint representations}

\subsection{Branching under $SU(n)\supset SU(3) \times SU(n-3) \times U(1)$ for the rank three antisymmetric tensor and adjoint
 representations}

We  assume that $SU(n)$ is broken by the ground state expectation of a single component $\overline{\phi}^{[123]}=a\neq 0$, corresponding to the simplest
case considered by Cummins and King \cite{cummins}, which applies for all $n$. The conditions on the scalar potential for
this case to apply will be given in Sec. 3.  Let us now divide the tensor indices into two classes,
\begin{align}\label{classes}
{\cal A}=&\{1,2,3\}~~~,\cr
{\cal B}=&\{4,...,n\}~~~.\cr
\end{align}
To get the needed branching expansions, we have to enumerate the possibilities for tensor indices to belong
to these two classes.  We use the notation $\big(R_{SU(3)}, R_{SU(n-3)}\big)(g)$, with $g$ the $U(1)$ generator eigenvalue.
Writing the $U(1)$ generator $G$  as
\begin{equation}\label{u1gen}
G={\rm Diag}(n-3,n-3,n-3,-3,-3,...,-3)
\end{equation}
with $n-3$ entries $-3$, the $U(1)$ generator value $g$ is simply
$n-3$ times the number of upper indices in ${\cal A}$ plus $-3$ times the
number of upper indices in ${\cal B}$; for lower indices the $U(1)$ contributions
are reversed in sign.  Since the overall normalization of the $U(1)$ generator
is arbitrary, normalization-independent statements  refer only to  {\it relative}
values of the $U(1)$ generators for different representations.

We begin by deriving the branching expansion for the $SU(n)$ rank three antisymmetric
tensor representation $n(n-1)(n-2)/6$, represented by the tensor $\phi^{[\alpha\beta\gamma]}$,
enumerating cases as follows.
\begin{enumerate}

\item{3 indices in ${\cal A}$.~~}
This corresponds to the representation $\big(1,1\big)(3n-9)$.

\item{2 indices in ${\cal A}$, 1 index in ${\cal B}$.~~}
Since the index in ${\cal B}$ can be chosen $n-3$ ways, this
corresponds to the representation $\big(\overline{3},n-3\big)(2n-9)$.

\item{1 index in ${\cal A}$, 2 indices in ${\cal B}$.~~}
Since the indices in ${\cal B}$ can be chosen $(n-3)(n-4)/2$ ways,
this corresponds to the representation $\big(3, (n-3)(n-4)/2\big)(n-9)$.

\item{3 indices in ${\cal B}$.~~}
Since the indices in ${\cal B}$ can be chosen $(n-3)(n-4)(n-5)/6$
ways, this corresponds to the representation $\big(1,(n-3)(n-4)(n-5)/6\big)(-9)$.
 \end{enumerate}

Thus we have the branching expansion, for $n>8$,
\begin{align}\label{branch1}
\frac{n(n-1)(n-2)}{6}=&\big(1,1\big)(3n-9)+\big(\overline{3},n-3\big)(2n-9)\cr
+&\left(3,\frac{(n-3)(n-4)}{2}\right)(n-9)
+\left(1,\frac{(n-3)(n-4)(n-5)}{6}\right)(-9)~~~.\cr
\end{align}
As a check on the counting, we note the identity
\begin{equation}\label{check1}
\frac{n(n-1)(n-2)}{6}=1+3(n-3)+3\frac{(n-3)(n-4)}{2}+\frac{(n-3)(n-4)(n-5)}{6}~~~.
\end{equation}
For the case $n=8$ discussed in \cite{adler1}, the $SU(5)$ three
upper index antisymmetric tensor is equivalent, by use of the
invariant tensor $\epsilon_{\alpha\beta\gamma\delta\epsilon}$, to
the $SU(5)$ two lower index antisymmetric tensor, and so represents
a $\overline{10}$ rather than a $10$. Thus we get the expansion
\begin{equation}\label{branch1a}
56=(1,1)(15)+(\overline{3},5)(7)+(3,10)(-1)+(1,\overline{10})(-9)~~~.
\end{equation}
This agrees with the expansion given in \cite{adler1} and the Slansky tables \cite{slansky},
apart from the fact that in this paper we have chosen the opposite sign convention for
the $U(1)$ generator $G$. For $n<8$, one makes similar conversions of upper index tensors to lower index ones  in Eq.
\eqref{branch1} ,  when the number of lower indices can be made smaller than the number of upper indices,
with the corresponding replacement of the representation $R$ by $\overline{R}$.  We also note that when
$n-3$ is divisible by 3, the $U(1)$ generator values can all be divided by 3, and this is the convention
that is used in the Slansky tables \big(see e.g. the expansion for the 20 of $SU(6)$\big).

We turn next to the branching expansion for the $SU(n)$ adjoint
 representation $n^2-1$, represented by the tensor $\phi^{\alpha}_{\beta}$, with
 $\sum_{\alpha}\phi^{\alpha}_{\alpha}=0$, again enumerating cases.

\begin{enumerate}

\item{diagonal traceless part analogous  to the $U(1)$ generator $G$.~~}
This corresponds to the representation $\big(1,1\big)(0)$.

\item{upper index and lower index both in ${\cal A}$, traceless part.~~}
This corresponds to the representation $\big(8,1\big)(0)$.

\item{upper index and lower index both in ${\cal B}$, traceless part.~~}
This corresponds to the representation $\big(1,(n-3)^2-1\big)(0)$.

\item{upper index in ${\cal A}$, lower index in ${\cal B}$.~~}
This corresponds to the representation $\big(3,\overline{n-3}\big)(n)$.

\item{lower index in ${\cal A}$, upper index in ${\cal B}$.~~}
This corresponds to the representation $\big(\overline{3},n-3\big)(-n)$.
\end{enumerate}

Thus we have the branching expansion
 \begin{equation}\label{branch2}
 n^2-1=\big(1,1\big)(0)+\big(8,1\big)(0)+\big(1,(n-3)^2-1\big)(0)+\big(3,\overline{n-3}\big)(n)+\big(\overline{3},n-3\big)(-n)~~~.
\end{equation}
As a check on the counting, we note the identity
\begin{equation}\label{check2}
n^2-1=1+8+(n-3)^2-1+6(n-3)~~~.
\end{equation}

We now note the phenomenon discussed in the $n=8$ case in \cite{adler1,adler2}, that the $U(1)$ generator
 of the $(\overline{3},n-3)$ is $-n$ in the branching expansion for the adjoint, whereas it
is $2n-9$ in the branching expansion for the rank three antisymmetric tensor.   The difference
between these two $U(1)$ generators is $2n-9-(-n)=3n-9=3(n-3)$, which is just the $U(1)$ generator
of the singlet $(1,1)$ in the expansion of Eq. \eqref{branch1}.  This is a direct result of the fact that
the $\overline{3}$ is represented by a two upper index antisymmetric tensor in the expansion of
Eq. \eqref{branch1}, and by a one lower index  tensor in the expansion of
Eq. \eqref{branch2}, so the difference in $U(1)$ generator values is $\big(2-(-1)\big)(n-3)=3(n-3)$.
When we discuss the scalar potential in Sec. 3, we will see that the complex states $(\overline{3},n-3)$ in
Eq. \eqref{branch1} are zero mass Goldstone modes. When the rank three antisymmetric tensor
is used to break the $SU(n)$ symmetry, the Goldstone modes are absorbed as longitudinal parts of the $(3,\overline{n-3})+(\overline{3},n-3)$
in the adjoint.  This is possible, even though the $U(1)$ generators do not match, because for the $(1,1)(3n-9)$
to get a ground state expectation value, the ground state must have a periodic structure modulo an integer divisor
of $3n-9$, and so the mismatch of the $U(1)$ generator values is equivalent to zero.

\subsection{Branching under $SU(n)\supset SU(2) \times SU(n-2) \times U(1)$ for the rank two antisymmetric tensor and adjoint
representations}

In this case we shall assume that $SU(n)$ is broken by the ground state expectation of a single component $\overline{\phi}^{[12]}=a\neq0$,
corresponding to the case studied by Li \cite{li}.  We now define the index classes by
\begin{align}\label{classes2}
{\cal A}=&\{1,2\}~~~,\cr
{\cal B}=&\{3,...,n\}~~~,\cr
\end{align}
and use the notation $(R_{SU(2)}, R_{SU(n-2)})(g)$, with $g$ the $U(1)$ generator.
Writing the $U(1)$ generator $G$  as
\begin{equation}\label{u1gen1}
G={\rm Diag}(n-2,n-2,-2,-2,...,-2)
\end{equation}
with $n-2$ entries $-2$, the $U(1)$ generator value $g$ is simply
$n-2$ times the number of upper indices in ${\cal A}$ plus $-2$ times the
number of upper indices in ${\cal B}$; for lower indices the $U(1)$ contributions
are reversed in sign.  Again, since the overall normalization of the $U(1)$ generator
is arbitrary, normalization-independent statements  refer only to  {\it relative}
values of the $U(1)$ generators for different representations.

Since the enumeration of cases parallels that in the rank three case, we go directly to the results.
For the rank two antisymmetric tensor, we have for $n>5$
\begin{equation}\label{branch3}
\frac{n(n-1)}{2}=\big(1,1\big)(2n-4)+ \big(2,n-2\big)(n-4)+\left(1,\frac{(n-2)(n-3)}{2}\right)(-4)~~~,
\end{equation}
with the three terms corresponding, respectively, to zero, one, and two upper
indices in ${\cal B}$.
As a check on the counting, we note the identity
\begin{equation}\label{check3}
\frac{n(n-1)}{2}=1+ 2(n-2)+\frac{(n-2)(n-3)}{2}~~~.
\end{equation}
For the case $n=5$, since the $SU(3)$ two upper index antisymmetric  tensor
represents a $\overline{3}$, we get the expansion
\begin{equation}\label{branch3a}
10=(1,1)(6)+(2,3)(1)+(1,\overline{3})(-4)~~~,
\end{equation}
in agreement with the expansion given in the Slansky tables \cite{slansky}.
When $n-2$ is divisible by 2, the $U(1)$ generator values can all be divided by
2, and this is the convention used in the Slansky tables \big(see, e.g., the expansion
for the 15 of $SU(6)$.\big)

For the adjoint representation $n^2-1$ of $SU(n)$, we get the branching expansion
\begin{equation}\label{branch4}
n^2-1=\big(1,1\big)(0)+\big(3,1\big)(0)+\big(1,(n-2)^2-1\big)(0)+\big(2,\overline{n-2}\big)(n)+\big(2,n-2\big)(-n)~~~,
\end{equation}
and as a check on counting
\begin{equation}\label{check4}
n^2-1=1+3+(n-2)^2-1+4(n-2)~~~.
\end{equation}

We again see the mismatch discussed in \cite{adler2} in the n=5 case.  The $U(1)$ generator
of the $(2,n-2)$ is $-n$ in the branching expansion of the adjoint, whereas it is $n-4$ in the
branching expansion for the rank two antisymmetric tensor.  The difference between these
two $U(1)$ values is $n-4-(-n)=2n-4=2(n-2)$, which is the $U(1)$ generator of the singlet
$(1,1)$ in the expansion of Eq. \eqref{branch3}.  This results from the fact that the $2$ is
represented by a one upper index tensor in the expansion of Eq. \eqref{branch3}, and by a one
lower index tensor in the expansion of Eq. \eqref{branch4}, with a resulting difference
of $U(1)$ generator values $\big(1-(-1)\big)(n-2)=2(n-2)$.  When we discuss the scalar potential in Sec. 3, we will see that the complex states $(2,n-2)$ in
Eq. \eqref{branch3} are zero mass Goldstone modes. When the rank two antisymmetric tensor
is used to break the $SU(n)$ symmetry, the Goldstone modes are absorbed as longitudinal parts of the $(2,\overline{n-2})+(2,n-2)$
in the adjoint.  This is possible, despite the $U(1)$ generator mismatch, because for the $(1,1)(2n-4)$
to get a ground state expectation value, the ground state must have a periodic structure modulo an integer divisor of $2n-4$, and so
the mismatch of the $U(1)$ generator values is equivalent to zero.

\section{Residual scalar masses}

In this section we analyze the residual scalar masses arising from
$SU(n)$ symmetry breaking with a general renormalizable scalar potential, first for a rank three
antisymmetric tensor scalar, and then for a rank two antisymmetric tensor.

\subsection{Residual scalar masses for $SU(n)$ symmetry breaking by a rank three antisymmetric tensor}

The most general $SU(n)$ invariant fourth degree potential formed from $\phi^{[\alpha\beta\gamma]}$, where the indices all range
from $1$ to $n$, has the form  \cite{cummins}
\begin{equation}\label{fourthorderpot}
V(\phi)=-\frac{1}{2}\mu^2\sum_{\alpha\beta\gamma} \phi^*_{[\alpha\beta\gamma]}\phi^{[\alpha\beta\gamma]}
+\frac{1}{4}\lambda_1\big(\sum_{\alpha\beta\gamma} \phi^*_{[\alpha\beta\gamma]}\phi^{[\alpha\beta\gamma]}\big)^2
+\frac{1}{4}\lambda_2 \sum_{\alpha\beta\gamma\rho\kappa\tau} \phi^*_{[\alpha \beta \gamma]}\phi^{[\alpha\beta\tau]}
\phi^*_{[\rho\kappa\tau]}\phi^{[\rho\kappa\gamma]}~~~.
\end{equation}
We assume $\mu^2>0$, so that the origin is a local maximum, and consider the case $\lambda_2<0$ studied in \cite{cummins}, for which   the potential  is bounded from below, for all $n$,  when
$3\lambda_1+\lambda_2>0$,
\begin{equation}\label{potbound}
V(\phi) \geq -\frac{3}{4} \frac{\mu^4}{3\lambda_1+ \lambda_2}~~~.
\end{equation}
This lower bound is attained when only one component of $\phi$ is nonzero, and as in our branching analysis we take
the nonvanishing component to be $\overline{\phi}^{[123]} = a \neq 0$, where
\begin{equation}\label{avalue}
|a|^2=\frac{1}{2}\frac{\mu^2}{3 \lambda_1+\lambda_2}~~~.
\end{equation}
We will derive Eqs. \eqref{potbound} and \eqref{avalue} shortly.

Continuing to follow \cite{cummins}, we  note that the potential of Eq. \eqref{fourthorderpot}
can be rewritten in terms of
\begin{equation}\label{eq:thetadef}
\theta_{\gamma}^{\tau}\equiv \sum_{\alpha\beta} \phi^*_{[\alpha \beta \gamma]}
\phi^{[\alpha\beta\tau]}~~~,
\end{equation}
which obeys $(\theta_{\gamma}^{\tau})^*=\theta_{\tau}^{\gamma}$, as
\begin{align}\label{fourthorderpot1}
V(\phi)=&-\frac{1}{2}\mu^2 \sum_{\gamma}\theta_{\gamma}^{\gamma} +
\frac{1}{4}\lambda_1 (\sum_{\gamma}\theta_{\gamma}^{\gamma})^2
+\frac{1}{4}\lambda_2\sum_{\gamma\tau}\theta_{\gamma}^{\tau}\theta_{\tau}^{\gamma}\cr
=&-\frac{1}{2}\mu^2 \sum_{\gamma}\theta_{\gamma}^{\gamma} +
\frac{1}{4}\lambda_1 (\sum_{\gamma}\theta_{\gamma}^{\gamma})^2
+\frac{1}{4}\lambda_2\sum_{\gamma}(\theta_{\gamma}^{\gamma})^2
+\frac{1}{2}\lambda_2\sum_{\gamma<\tau}\theta_{\gamma}^{\tau}\theta_{\tau}^{\gamma}~~~.\cr
\end{align}

To expand the potential around its minimum, we substitute
\begin{equation}\label{sigdef}
\phi^{[\alpha\beta\gamma]}=\overline{\phi}^{[\alpha\beta\gamma]}+ \sigma^{[\alpha\beta\gamma]}~~~,
\end{equation}
where $\overline{\phi}^{[\alpha\beta\gamma]}=a\epsilon^{\alpha\beta\gamma}$ is nonzero only when its tensor indices are some permutation of
$1,2,3$.
For $\theta_{\gamma}^{\tau}$ we find
\begin{equation}\label{thetacomp}
\theta_{\gamma}^{\tau}=2\sum_{\alpha<\beta}\big(\overline{\phi}^{\,*}_{[\alpha\beta\gamma]}\overline{\phi}^{[\alpha\beta\tau]}+
\overline{\phi}^{\,*}_{[\alpha\beta\gamma]}\sigma^{[\alpha\beta\tau]}+\sigma^*_{[\alpha\beta\gamma]}\overline{\phi}^{[\alpha\beta\tau]}
+\sigma^*_{[\alpha\beta\gamma]}\sigma^{[\alpha\beta\tau]}\big)~~~.
\end{equation}

We consider first the term $\sum_{\gamma<\tau}\theta_{\gamma}^{\tau}\theta_{\tau}^{\gamma}$ in Eq. \eqref{fourthorderpot1}.
The term in  $\theta_{\gamma}^{\tau}$ that is quadratic in $\overline{\phi}$ must have $\gamma=\tau$, and so does not
contribute to this sum over $\gamma<\tau$.  Hence the term in  $\theta_{\gamma}^{\tau}$   that is quadratic in $\sigma$ makes
a contribution to  this sum that is
third order in $\sigma$, and can be dropped in calculating the potential to second order in $\sigma$.  Thus we get
\begin{align}\label{summ1}
\sum_{\gamma<\tau}\theta_{\gamma}^{\tau}\theta_{\tau}^{\gamma}=& \sum_{\gamma<\tau}|\theta_{\gamma}^{\tau}|^2\cr
=&4\sum_{\gamma<\tau}|\sum_{\alpha<\beta}\big(\overline{\phi}^{\,*}_{[\alpha\beta\gamma]}\sigma^{[\alpha\beta\tau]}
+\sigma^*_{[\alpha\beta\gamma]}\overline{\phi}^{[\alpha\beta\tau]}\big)|^2\cr
=&4\sum_{\gamma<\tau}|\sum_{\alpha<\beta}\overline{\phi}^{\,*}_{[\alpha\beta\gamma]}\sigma^{[\alpha\beta\tau]}|^2~~~,\cr
\end{align}
where in the final line we have used the fact that when $\alpha,\, \beta,\, \tau$ are permutations of $1,2,3$,
then when $\tau=1$ there is no $\gamma$ obeying $\gamma<\tau$, and when $\tau=2$ or $\tau=3$, the $\gamma$ obeying
$\gamma<\tau$ must equal either $\alpha$ or $\beta$, and so the factor $\sigma^*_{[\alpha\beta\gamma]}$ multiplying
$\overline{\phi}^{[\alpha\beta\tau]}$  vanishes.  By similar reasoning, the sum over $\tau$ in the final
line of Eq. \eqref{summ1} must range from $4$ to $n$ independent of the values of $\alpha<\beta$, since if $\tau\leq 3$, then $\gamma \leq 2$ and either the first or the second factor vanishes.  Hence we get
\begin{equation}\label{summ2}
\sum_{\gamma<\tau}\theta_{\gamma}^{\tau}\theta_{\tau}^{\gamma}=4|a|^2 \sum_{(\alpha,\,\beta)=(1,2),\,(1,3),\,(2,3)}~
\sum_{\tau=4}^n|\sigma^{[\alpha\beta\tau]}|^2~~~.
\end{equation}
Since $\sigma^{[\alpha\beta\tau]}$ in this equation has $\alpha \in {\cal A}$, $\beta \in {\cal A}$, and $\tau \in {\cal B}$,
it belongs to the representation $(\overline{3},n-3)$, and so we can rewrite Eq. \eqref{summ2} as
\begin{equation}\label{summ3}
\sum_{\gamma<\tau}\theta_{\gamma}^{\tau}\theta_{\tau}^{\gamma}=4|a|^2 \sum_{k=1}^3 \sum_{l=1}^{n-3}
|\sigma(\overline{3},k; n-3,l)|^2~~~.
\end{equation}

The remaining terms in  Eq. \eqref{fourthorderpot1} all involve the diagonal element $\theta_{\gamma}^{\gamma}$,
which from Eq. \eqref{thetacomp} is given by
\begin{equation}\label{thetacomp1}
\theta_{\gamma}^{\gamma}=2\sum_{\alpha<\beta}\big(|\overline{\phi}^{[\alpha\beta\gamma]}|^2
+2 {\rm Re}\big(\overline{\phi}^{\,*}_{[\alpha\beta\gamma]}\sigma^{[\alpha\beta\gamma]}\big)
+|\sigma^{[\alpha\beta\gamma]}|^2\big)~~~.
\end{equation}
From this,  substituting $\overline{\phi}^{[\alpha\beta\gamma]}=a\epsilon^{\alpha\beta\gamma}$, splitting sums on $\gamma$ into
 disjoint sums $\sum_{\gamma \in {\cal A}}$ and $\sum_{\gamma \in {\cal B}}$, and dropping terms of higher order
than quadratic in $\sigma$, one finds
\begin{align}\label{thetasums}
\sum_{\gamma}(\theta_{\gamma}^{\gamma})^2=&12\Big(|a|^4+4|a|^2 {\rm Re}(a^*\sigma^{[123]}) +2|a|^2|\sigma^{[123]}|^2
+4 \big( {\rm Re}(a^*\sigma^{[123]})\big)^2\Big)~~~,\cr
\sum_{\gamma}\theta_{\gamma}^{\gamma}=&6\Big(|a|^2 +2{\rm Re}(a^*\sigma^{[123]})+|\sigma^{[123]}|^2 \Big)+
 \sum_{\alpha\beta}\sum_{\gamma\in {\cal B}} |\sigma^{[\alpha\beta\gamma]}|^2~~~,\cr
 \Big(\sum_{\gamma}\theta_{\gamma}^{\gamma}\Big)^2=&36\Big(|a|^4+4|a|^2 {\rm Re}(a^*\sigma^{[123]}) +2|a|^2|\sigma^{[123]}|^2
+4 \big( {\rm Re}(a^*\sigma^{[123]})\big)^2\Big)\cr
+&12|a|^2 \sum_{\alpha\beta}\sum_{\gamma\in {\cal B}} |\sigma^{[\alpha\beta\gamma]}|^2~~~.\cr
\end{align}

Substituting Eqs. \eqref{summ3} and \eqref{thetasums} into Eq. \eqref{fourthorderpot1}, and combining the first order terms in
$\sigma$, we get
\begin{equation}\label{firstorder}
{\rm Re}(a^*\sigma^{[123]})(-6\mu^2+36\lambda_1|a|^2+12 \lambda_2 |a|^2)~~~,
\end{equation}
which when equated to zero gives Eq. \eqref{avalue}.  Using this value of $|a|^2$, we find the lower bound of Eq. \eqref{potbound}
for the value of the potential at the minimum.  Splitting the
sum $\sum_{\alpha\beta}\sum_{\gamma\in {\cal B}} |\sigma^{[\alpha\beta\gamma]}|^2$ into three pieces,
\begin{equation}\label{sumsplit}
\sum_{\alpha\beta}\sum_{\gamma\in {\cal B}} |\sigma^{[\alpha\beta\gamma]}|^2
=\big(\sum_{\alpha\beta \in {\cal A}} ~ \sum_{\gamma\in {\cal B}}
+ 2 \sum_{\alpha \in {\cal A}}~\sum_{\beta\gamma \in {\cal B}}
+\sum_{\alpha\beta\gamma \in {\cal B}}\big)|\sigma^{[\alpha\beta\gamma]}|^2~~~,
\end{equation}
and relabeling $\sigma^{[\alpha\beta\gamma]}$ in terms of the representations
appearing in the branching expansion of Eq. \eqref{branch1}, we get as the final result for the
expansion of the potential near the minimum through second order terms,
\begin{align}\label{finalresult}
V(\overline{\phi}+\sigma)=& -\frac{3}{4}\frac{\mu^4}{3\lambda_1+\lambda_2}\cr
+&\left( {\rm Re}(\frac{a^*}{|a|}\sigma(1,1)) \right)^2 6 \mu^2 \cr
+&\sum_{k=1}^3 \sum_{l=1}^{n-3}|\sigma(\overline{3},k; n-3,l)|^2 \times  0  \cr
+&\sum_{k=1}^3 \sum_{l=1}^{(n-3)(n-4)/2}  |\sigma(3,k; (n-3)(n-4)/2,l)|^2 2 \mu^2 \frac{-\lambda_2}{3\lambda_1+\lambda_2}\cr
+&\sum_{l=1}^{(n-3)(n-4)(n-5)/6}  |\sigma(1; (n-3)(n-4)(n-5)/6,l)|^2  3 \mu^2 \frac{-\lambda_2}{3\lambda_1+\lambda_2}~~~.\cr
\end{align}
The remarks made in Sec. 2 about using the rank $n-3$ epsilon tensor to
replace upper index tensors by lower index tensors in conjugate representations, when this
reduces the number of indices, applies here. We see that as noted in Sec. 2, the Goldstone modes, with mass $0$, are
in the representation $(\overline{3},n-3)$, which has a $U(1)$ generator mismatch with respect to the corresponding
representation in the expansion of the adjoint representation.

\subsection{Residual scalar masses for $SU(n)$ symmetry breaking by a rank two antisymmetric tensor}

The most general $SU(n)$ invariant fourth degree potential formed from the rank two antisymmetric tensor scalar $\phi^{[\alpha\beta]}$, where the indices all range from $1$ to $n$, has the form \cite{li} for $n>4$,\footnote{A perceptive referee has pointed out that for $SU(4)$ there is an exception; one can construct the invariant 
 $\phi^{\alpha\beta}\phi^{\gamma\delta}\epsilon_{\alpha\beta\gamma\delta}+{\rm adjoint}$, and so the most 
general renormalizable potential has a more complicated form than Eq. \eqref{fourthorderpot2}.  For rank three antisymmetric tensors in $SU(6)$ the analog of this 
invariant vanishes by antisymmetry of the epsilon tensor, so there is not a similar exception to the potential of Eq. \eqref{fourthorderpot}.  The paper of Li 
\cite{li} overlooked the rank two exception because it first treated rank two symmetric tensors, and then took the same potential for the antisymmetric 
tensor case.}
\begin{equation}\label{fourthorderpot2}
V(\phi)=-\frac{1}{2}\mu^2\sum_{\alpha\beta} \phi^*_{[\alpha\beta]}\phi^{[\alpha\beta]}
+\frac{1}{4}\lambda_1\big(\sum_{\alpha\beta} \phi^*_{[\alpha\beta]}\phi^{[\alpha\beta]}\big)^2
+\frac{1}{4}\lambda_2\sum_{\alpha\beta\rho\tau} \phi^*_{[\alpha \rho ]}\phi^{[\alpha\tau]}
\phi^*_{[\beta\tau]}\phi^{[\beta\rho]}~~~.
\end{equation}
Since the method of analysis parallels that used in the rank three case, we state only the final results.
We assume that $\lambda_2<0$ and $2\lambda_1 + \lambda_2 >0$, and as
in our branching analysis we take
the nonvanishing component of $\phi^{[\alpha\beta]}$ to be $\overline{\phi}^{[12]} = a \neq 0$.
The potential minimum is at
\begin{equation}\label{avalue1}
|a|^2=\frac{\mu^2}{2 \lambda_1+\lambda_2}~~~,
\end{equation}
and the value of the potential at the minimum is
\begin{equation}\label{potvalue1}
 -\frac{1}{2} \frac{\mu^4}{2\lambda_1+ \lambda_2}~~~.
\end{equation}
For the expansion of the potential through second order terms, we find
\begin{align}\label{finalresult1}
V(\overline{\phi}+\sigma)=& -\frac{1}{2}\frac{\mu^4}{2\lambda_1+\lambda_2}\cr
+&\left( {\rm Re}(\frac{a^*}{|a|}\sigma(1,1)) \right)^2 2 \mu^2 \cr
+&\sum_{k=1}^2 \sum_{l=1}^{n-2}|\sigma(2,k; n-2,l)|^2 \times  0  \cr
+&\sum_{l=1}^{(n-2)(n-3)/2}  |\sigma(1; (n-2)(n-3)/2,l)|^2   \mu^2 \frac{-\lambda_2}{2\lambda_1+\lambda_2}~~~.\cr
\end{align}
The remarks made in Sec. 2 about using the rank $n-2$ epsilon tensor to
replace upper index tensors by lower index tensors in conjugate representations, when this
reduces the number of indices, applies here. As noted in Sec. 2, the zero mass Goldstone modes are
in the representation $(2,n-2)$, which has a $U(1)$ generator mismatch with respect to the corresponding
representation in the expansion of the adjoint representation.

\section{Summary}

We have derived further properties of $SU(n)$  symmetry breaking by rank three and rank two antisymmetric
tensor scalars, extending previous analyses in the literature.  The $U(1)$ generator mismatch
highlighted in \cite{adler1}, \cite{adler2} is seen to originate from the fact that the $SU(3)$ representation
$\overline{3}$  can be represented by a two upper index antisymmetric tensor, or a one lower index tensor, the
former occurring in the branching expansion for the rank three antisymmetric tensor, and the latter in the
branching expansion for the adjoint.  An analogous statement holds for the $SU(2)$ representation $\overline{2}\equiv 2$
in the rank two antisymmetric tensor case.  The results of Eqs. \eqref{finalresult} and \eqref{finalresult1} for
residual scalar masses will be of use in model building in which $SU(n)$ symmetry is broken by a rank three or rank two
antisymmetric tensor scalar field.

\end{document}